\documentclass[a4paper,aps,twocolumn,prb,showpacs]{revtex4}
\usepackage[dvips]{color}
\usepackage{fancyhdr}
\usepackage{graphicx}
\usepackage{lscape}
\usepackage{amsmath}
\usepackage{amssymb}
\usepackage{mathrsfs}
\usepackage{bm}
\usepackage{epsfig}
\usepackage{dcolumn}
\usepackage[english]{babel}
\newcommand{\be}{\begin{equation}}
\newcommand{\ee}{\end{equation}}
\newcommand{\beq}{\begin{eqnarray}}
\newcommand{\eeq}{\end{eqnarray}}
\newcommand{\ba}{\begin{array}}
\newcommand{\ea}{\end{array}}
\newcommand{\bea}{\begin{eqnarray}}
\newcommand{\eea}{\end{eqnarray}}
\newcommand{\ex}[1]{\mbox{e}^{#1}}

\newcommand{\bra}[1]{| #1 \rangle}

\begin{document}

\title{The quantum optical Josephson interferometer}

\author{Dario Gerace,$^{1,2}$ Hakan E. T\"ureci,$^1$  A. Imamo\v{g}lu,$^1$ 
Vittorio Giovannetti$^3$ and Rosario Fazio$^{3,4}$}
\affiliation{ $^1$Institute of Quantum Electronics, ETH Zurich, 8093 Zurich
(Switzerland) \\
$^2$CNISM and Dipartimento di Fisica ``A. Volta,'' Universit\`a di Pavia, 27100 Pavia (Italy)\\
$^3$NEST (CNR-INFM) and Scuola Normale Superiore,
Piazza dei Cavalieri 7, 56126 Pisa (Italy) \\
$^4$International School for Advanced Studies (SISSA),
via Beirut $2-4$,  34014 Trieste (Italy) }


\begin{abstract}
The interplay between coherent tunnel coupling and on-site interactions in dissipation-free bosonic
systems has lead to many spectacular observations, ranging from 
the demonstration of number-phase uncertainty relation to quantum phase transitions. 
To explore the effect of dissipation and coherent drive on tunnel coupled
interacting bosonic systems, we propose a device that is the quantum optical analog of a 
Josephson interferometer. It consists of two coherently driven linear
optical cavities connected via a central cavity with a single-photon nonlinearity. The Josephson-like
oscillations in the light emitted from the central cavity as a function of the phase difference
between two pumping fields can be suppressed by increasing the strength of the nonlinear coupling.
Remarkably, we find that in the limit of ultra-strong interactions in the center-cavity, the coupled
system maps on to an effective Jaynes-Cummings system with a nonlinearity determined by the tunnel
coupling strength. In the limit of a single nonlinear cavity coupled to two linear waveguides, the
degree of photon antibunching from the nonlinear cavity provides an excellent measure of the
transition to the nonlinear regime where Josephson oscillations are suppressed.
\end{abstract}


\maketitle

\section{Introduction}
Cavity quantum electrodynamics (QED) experiments based on strong coupling of a single anharmonic emitter to a cavity-mode have lead to the observation of the photon blockade effect where photon-photon interactions alter the statistics of light emitted by the cavity. Experimental results showing nonlinearities at the single-photon level have recently been achieved in both atomic\cite{birnbaum05nat,schuster08np} and solid-state QED systems\cite{schuster07nat,kevin07nat,kartik07nat,jelena08} with single cavities. Motivated by the success of single-cavity QED experiments, center of attention has now shifted to the exploration of the rich physics promised by strongly-correlated quantum optical systems in multi-cavity and extended photonic media. Several works have recently considered this possibility: a photonic version of the Bose-Hubbard model with an array of nonlinear cavities\cite{hartmann06,greentree06,angelakis07}, the realization of the Tonks-Girardeau regime of interacting bosons in a non-linear optical fiber\cite{chang08}, and a photonic analogue of the Kondo effect in a 1D waveguide\cite{shenF07}. Most of the existing proposals have focused so far on strongly-correlated photonic systems in the quasi-equilibrium regime for which dissipation is essentially negligible. In most realistic cavity-QED structures on the other hand, dissipation does play a substantial role and it is very difficult to fix the photon number. When photon losses cannot be neglected, the system reaches a stationary state given by the balance of dissipation and driving. We propose here an optical analogue of the superconducting Josephson interferometer, which we name the \textit{quantum optical Josephson interferometer}, revealing new features due to the genuine non-equilibrium interplay of coherent tunneling and on-site interactions.

\begin{figure}[t]
\begin{center}
\includegraphics[width=0.45\textwidth]{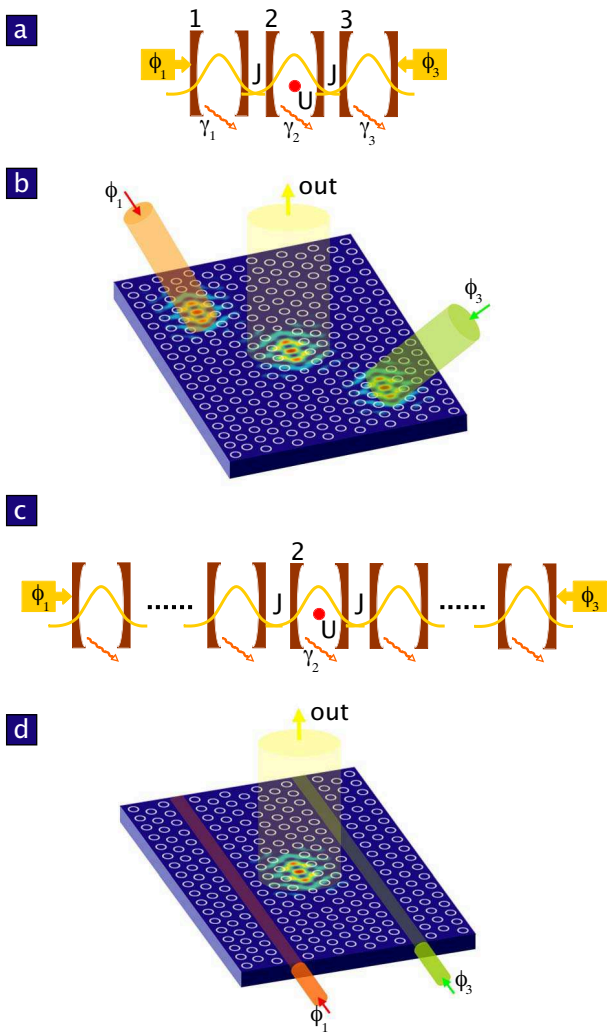}
\caption{The systems under consideration.
(a) Schematic of the quantum optical Josephson interferometer with three coupled cavities. The relevant quantities of the model are also defined. (b) A possible photonic crystal-based implementation (with calculated cavity mode profiles), where the middle cavity can contain a quantum dot or a quantum well in strong coupling with the high-Q photonic crystal cavity mode. (c) Schematic of an interferometer in the limit of a large number of coupled linear cavities in optical contact with the central nonlinear cavity, in which only the edges of the system are pumped. (d) A possible solid-state implementation of (c)
employing a photonic crystal circuit with two side-waveguides coupled to the central nonlinear cavity.
}\label{fig1}
\end{center}
\end{figure}

The two variants of the device we are presenting couple a central nonlinear cavity to two external driving lasers through either two side cavities (Fig. 1a,b) or two waveguides (Fig. 1d). The three-cavity system can be generalized to an N-cavity system with the central nonlinear one\cite{nori08} (Fig. 1c), and in the limiting case of very large N this reduces to the single cavity coupled to two side-waveguides (Fig. 1d) .
In both cases, the coupling to the side cavities (or waveguides) is a consequence of photon tunneling. We assume the center-cavity can be tuned to have a sizable single-photon nonlinearity, e.g. due to some radiation-matter interaction effect, be it Jaynes-Cummings-type dynamics  (with a single atom or quantum dot in the center cavity)~\cite{jc}, giant Kerr nonlinearity~\cite{imamoglu97prl}, or 0D polariton interaction (e.g., with a quantum well embedded in the center-cavity)~\cite{ciuti06prb}. The model discussed here is quite general and can be realized in a variety of quantum optical systems. In the following we show that light emitted from the center cavity is the result of two competing effects, tunneling and interactions, leading to a crossover between the coherent and strongly correlated regimes. In the coherent regime, photons are delocalized over the three cavities and the emitted light is strongly dependent on the phase difference between the two pumping lasers (Josephson oscillations). In the strongly correlated regime, the inhibition of photon number occupation beyond Fock states $|0\rangle$ and $|1\rangle$ in the center cavity reduces the quantum coherence between the two outer ones. The suppression of Josephson oscillations in the emitted light is accompanied by a crossover from Poissonian to sub-Poissonian photon statistics. 

The main thrust of the present work is to investigate the crossover from coherent to correlated regimes of this intrinsically non-equilibrium system by detecting light emitted from the center cavity. We show that photon correlation measurements reveal features of an interacting few-body system that are not captured by more traditional transport-type measurements. 
In fact, the device we propose has close analogies with a phase-biased Cooper pair transistor. In the latter, the critical current can be electrostatically modulated by changing the gate potential on the central island connected to two superconducting reservoirs\cite{averin_likarev, matveev93prl}. The interplay between coherent tunnel coupling and on-site interactions in these systems has been utilized in many ground-breaking experiments, ranging from the observation of a quantum phase transition in Josephson junction arrays\cite{geerligs89} to the direct demonstration of number-phase uncertainty in superconducting islands\cite{elion94}. While the operation of both electronic and photonic versions is based on the quantum mechanical conjugation between phase and number variables, the quantum optical Josephson interferometer is an intrinsically open system. Given the enormous impact of the Josephson devices in electronics, from metrology to quantum information processing, we expect that the quantum optical Josephson interferometer might constitute the building block for a new class of quantum optical devices.

\begin{figure*}[t]
\begin{center}
\includegraphics[width=0.8\textwidth]{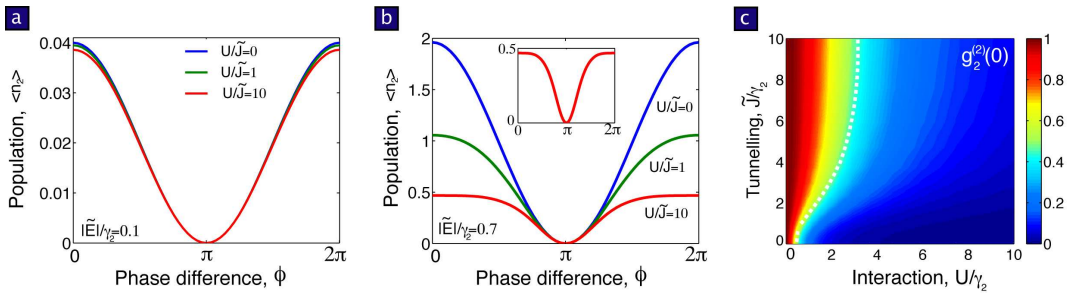}
\caption{Numerical solutions for the three cavity system.
(a) Average population in the center cavity as a function of $\phi$ for different values of $U/ \tilde{J}$,
in the weak pumping regime ($|\tilde{E}|/ \gamma_2=0.1$): Josephson oscillations are barely modified as $U$ is increased. (b)  The same quantity as in (a), but in a stronger pumping regime
($|\tilde{E}|/ \gamma_2=0.7$): Josephson oscillations are suppressed by increasing 
the interaction strength. The inset shows the curve calculated for $U/ \tilde{J}=10$.
(c)  Second-order correlation function as a function of $\tilde{J}$ and $U$ for $\gamma / \gamma_2=5$ 
under weak pumping conditions, $|\tilde{E}|/ \gamma_2=0.1$,
showing the transition from Poissonian (red) to sub-Poissonian (blue) light statistics. The functional dependence $U_{th}(\tilde{J})$ is highlighted with a dashed white line.} \label{fig2}
\end{center}
\end{figure*}

\section{Theoretical discussion}
In the absence of losses the three cavity setup depicted in Fig.~1a,b is described by the 
Hamiltonian
\begin{eqnarray}
\label{hamilton}
      \hat{H} =
      \sum_{k=1}^{3}\Delta_{k} \hat{p}_{k}^{\dagger}\hat{p}_{k}  &+&
      J(\hat{p}_1^{\dagger}\hat{p}_2 +
      \hat{p}_2^{\dagger}\hat{p}_3 + \mathrm{h.c.}) +
      U \hat{p}_2^{\dagger}\hat{p}_2^{\dagger}\hat{p}_2 \hat{p}_2 \nonumber \\
      &+& \sum_{k=1,3} (E_k \hat{p}_k^{\dagger} + \mathrm{h.c.})  \, ,
\end{eqnarray}
written in the rotating frame with respect to the frequencies  of the two pumping lasers ($\hbar=1$).
In the above equation, $\Delta_k=\omega_k - \omega_{\mathrm{L}}$  are the detunings of the coherent
pump lasers whose amplitudes are $E_{1,3}=|E_{1,3}| \exp\{i\phi_{1,3}\}$, 
where $|E_{1,3} | $ are
assumed time-independent (cw pumping). We will characterize the  three cavity system by analyzing the
response of the center-cavity as a function of the phase difference $\phi=\phi_3-\phi_1$. The two
couplings $J$ and $U$ quantify the hopping strength between neighboring cavities and the nonlinear
photon coupling in the center-cavity, respectively.
The operators $\hat{p}_{1,3}$  describe noninteracting bosonic fields in the 
external cavities, i.e. free cavity photons, while the elementary excitations in the center-cavity are interacting polaritons denoted by $\hat{p}_{2}$ . Both the tunneling and the coherent pumping act on purely photonic degrees of freedom, e.g. $J$ is due to the overlap of the photonic part among nearest neighbors cavities. A rigorous derivation of the model in Eq. (\ref{hamilton}) inevitably depends on the  specific system under consideration, and has been provided before in the context of atomic\cite{hartmann06} or solid-state\cite{ciuti06prb} cavity-QED implementations. Readily accessible schemes for achieving large photon-photon interactions are based on resonant coupling of a single emitter to a cavity mode (i.e. the Jaynes-Cummings model). As we detail in the Supplementary Information section (``Experimental Feasibility''), the principal features we obtain using the model (\ref{hamilton}) are qualitatively identical to those predicted by the Jaynes-Cummings model-type single-photon nonlinearity. 

The dynamics of the full model, Eq. (\ref{hamilton}),  is  effectively equivalent to that of two
coupled bosonic fields: one is coherently driven, while the other is nonlinear (see Methods section).
From now on we consider for simplicity the case of equal detunings and resonant pumping, $\Delta_k=\Delta=0$. Losses can be
taken into account within the quantum Master equation in Born-Markov approximation for the system
density matrix $\rho$, which is expressed in the usual Lindblad form~\cite{carmichael_book}. The
relevant Master equation for this model is
\begin{equation}
\label{master}
      \frac{\partial\rho}{\partial t} = i [\rho,\hat{H}]
      + \sum_{k=1}^{3}\frac{\gamma_k}{2}
      (2\hat{p}_k \rho\hat{p}_k^{\dagger}-\hat{p}_k^{\dagger}\hat{p}_k \rho
      - \rho \hat{p}_k^{\dagger}\hat{p}_k)\, .
\end{equation}
In most of the relevant regimes,  the Master equation has to be solved numerically. A description of
the approach used in this work is presented in the Methods section. In the remaining of the text, 
we will assume $\gamma_{1,3}=\gamma$. 
With the specific experimental settings of Figs. 1b and d in mind, inter-cavity tunnel coupling $J\simeq 1$ meV \cite{kapon08oe}, cavity quality factor $Q\simeq 10^5$, i.e. $\gamma\simeq 0.01$ meV in the optical/near-infrared domain\cite{derossi08apl}, and nonlinear to dissipation rate ratio $U/ \gamma =10$ (see Supplementary Information) can be realistically achieved, which makes the following theoretical analysis experimentally relevant.

It is instructive to first consider the case in which there is no interaction ($U=0$),  where an
exact analytical solution for the steady state of Eq.~(\ref{master}) can be obtained. The case of
equal amplitudes of the two driving lasers ($E_1=E_3 =  E $) and
equal losses in the three cavities ($\gamma=\gamma_2$) captures all the essential details of the
non-interacting case. In steady state the average number of photons in the central cavity $\langle
n_{2} \rangle= \langle \hat{p}_{2}^{\dagger}\hat{p}_{2} \rangle$ is found to be
\begin{equation}
\label{solutionUeq0}
  \langle n_2\rangle = \frac{64 J^2 | E |^2}{(8 J^2+\gamma^2)^2} \cos^2\frac{\phi}{2}\, .
\end{equation}
This is an analog of Josephson oscillations,  imprinted in the light emitted from the center-cavity,
due to the interference between the two coherent driving fields. Two features of this solution are to
be noticed for a comparison with the more interesting $U \neq 0$ situation treated below. First, the
size of the oscillations is maximized at $J \sim \gamma/2$, as a result of an interplay of dissipation
and interference. Moreover, while $\langle n_2(\phi=0)\rangle$ is suppressed and eventually goes to
zero for $J \gg \gamma $, the oscillations keep a cosine-like behavior as a function of $\phi$.

In Fig.~2, we present our numerical 
results for experimentally accessible  observables of the system when the
interaction is switched on ($U > 0$). Rescaled quantities $\tilde{J}$ and $\tilde{E}$ are defined 
for the effective two-cavity model as outlined in the Methods section. 
Do the Josephson oscillations in $\langle n_{2} (\phi) \rangle$, as measured by
detecting the light emitted from the center cavity (Fig.~1b), remain intact? In Fig.~2a, we plot
$\langle n_{2} (\phi) \rangle$ for various values of the interaction at a pumping strength of $|\tilde{E}|/ \gamma_2=0.1$. The size as well as the functional form of the oscillations barely
change as $U/\tilde{J}$ is varied across a wide range of values, under these weak pumping 
conditions. 
This picture changes dramatically when we pump the system stronger, shown in 
Fig.~2b for $|\tilde{E}|/\gamma=0.7$. 
Here, the average population in the center-cavity can be sizeable and nonlinear effects
are more pronounced. In contrast to the weak pumping case (Fig.~ 2a), the size of the oscillations is
suppressed to a great extent as $U$ is increased. Besides the strong suppression of visibility,
Fig.~2b shows a dramatic deviation from the cosine-like functional form, Eq.~(\ref{solutionUeq0}), as
$U/\tilde{J}$ is increased from zero (the behavior for  $U/\tilde{J}\gg 1$ is shown in the inset).

\begin{figure}[t]
\begin{center}
\includegraphics[width=0.45\textwidth]{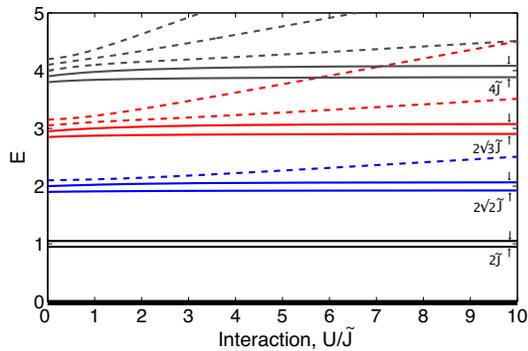}
\caption{Variation of the energy levels of the three-cavity Josephson interferometer as a function of 
$U/\tilde{J}$. We plot the first few total photon number manifolds in the range $N_{tot}=1$ to $N_{tot}=4$, and $\tilde{J}/\gamma_2=0.05$. The various $N_{tot}$ manifolds are marked on the vertical axis and set off from each other by an arbitrary $\omega_{\mathrm{L}}$ for visibility. The energy levels in each manifold undergo an anticrossing at $U/\tilde{J} \sim 1$ and a cross-over takes place to an effective Jaynes-Cummings sequence as $U/\tilde{J} \rightarrow \infty$.} \label{fig3}
\end{center}
\end{figure}

Next we investigate how this cross-over is reflected in the photon statistics of light emitted from
the center cavity. For this, in Fig.~2c we plot the zero-time delay second-order correlation function
$g^{(2)}_{2}(0)$  (See Methods section) as
a function of the scaled quantities $U/\gamma_2$ and $\tilde{J}/\gamma_2$.
We find that $g^{(2)}_{2}(0)$ displays a sharp transition from Poissonian to sub-Poissonian light
statistics as the interaction strength $U$ is increased. The threshold for anti-bunched
(sub-Poissonian) light generation, $U_{th}$, is a function of $J$. For $\tilde{J}/\gamma_2 \ll 1$, the
anti-bunching threshold is $U_{th} (\tilde{J}) \sim \gamma_2$, while for $J/\gamma_2 \gg 1$,  
$U_{th} (\tilde{J})
\sim \gamma + \gamma_2$. These two regimes are connected by a smooth crossover region. This peculiar
behavior of the anti-bunching threshold is related to the effective dissipation rates of the coupled
system as the coupling strength $\tilde{J}$ is varied. At small $\tilde{J}$, 
the coupling to the center cavity is
perturbative; the nonlinearity  (i.e. antibunching) therefore sets in when $U$ is larger than the
broadening of the bare center cavity polariton states i.e. $\gamma_2$. As $\tilde{J}$ is 
increased, the
coupling becomes non-perturbative and the relevant eigenstates of the coupled system are
superpositions of center and outer cavity states; such dressed states have broadening contributions
coming from both center and outer cavities and therefore the nonlinearity now has to be larger than
the broadening of the dressed states in order for the system to exhibit antibunching. A remarkably
simple expression for $g_2^{(2)}(0)$ can be derived in the weak pumping limit (See Discussion in
Supplementary Information section) which captures all the regimes discussed:
\begin{equation}
g^{(2)}_{2}(0) = \frac{\Gamma^2}{\Gamma^2 + 4\alpha^2(\tilde{J})U^2} \, ,
\end{equation}
where $\Gamma = \gamma + \gamma_2$ and $\alpha(\tilde{J}) = 
(4\tilde{J}^2 + \gamma\Gamma)/(4\tilde{J}^2 + \gamma\gamma_2)$.
These results  are consistent with the expectation that strong photon nonlinearity can lead to
photon-blockade \cite{imamoglu97prl} in the center cavity giving rise to anti-bunching. While the
{\em relative} strength of $U$ with respect to tunnel coupling $\tilde{J}$ 
seems to matter at small couplings $\tilde{J}/ \gamma_2$ (i.e. $U_{th}(\tilde{J})$ is a  
monotonic function of $\tilde{J}$), at larger $\tilde{J}$ the relative effect of $U$
saturates.

An interesting feature that does not leave any footprint in these two observables considered is the
nature of the system's effective non-linearity. For $U \ll \tilde{J}$, the 
deviation of the system energy levels from a harmonic structure is linearly proportional to $U$, and
this determines the main behaviour of $g^{(2)}_2(0)$. For $U \gg \tilde{J}$ 
however,  the center cavity acts as a two-level system (only
$0$ and $1$ photon states available), which is coupled to the linear cavities with strength $\tilde{J}$.
Thus, the coupled system maps onto an effective Jaynes-Cummings (JC) model where, surprisingly, the tunnel coupling strength plays the role that is commonly played by the atom-cavity dipole-coupling in the original JC model~\cite{jc}. In Fig.~3, we show how this comes about: 
two levels of each constant photon-number manifold ($N_{tot}=n_2 + n_s$) split off from the rest of the levels to form a JC-sequence as $U$ is increased beyond $\tilde{J}$.

\begin{figure*}[t]
\begin{center}
\includegraphics[width=0.8\textwidth]{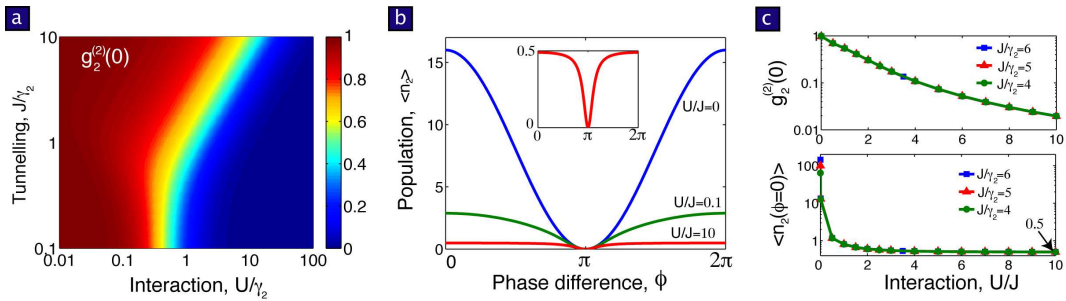}
\caption{Numerical solutions for the waveguide-coupled limit. We assume the coherent states 
$|\langle p_{1,3}\rangle|=1$. 
(a) Second-order correlation function at zero-time delay for light emitted from the center cavity, as a function of $U$ and $J$. A sharp crossover between Poissonian (red) and sub-Poissonian (blue) statistics is seen in the $U-J$ plane. (b) Average population in the center cavity for $J/ \gamma_2=2$ as a function of $\phi$, for different values of $U/J$. The oscillations approach the limiting case of Eq.~(6) when $U/J \gg 1$. The inset is a zoom on the curve for $U/ J =10$, showing $\langle n_2(\phi=0)\rangle\to 0.5$ and a strong deviation from simple cosine-like behaviour.
(c) Scaling of $g_2^{(2)}(0)$ and $\langle n_2 (\phi=0)\rangle$ as a function of $U/J$ for 
different values of $J/ \gamma_2$, showing a smooth crossover from delocalized to
localized regime for $U/J > 1$.} \label{fig4}
\end{center}
\end{figure*}

In the limit of an infinite number of linear cavities  coupled on either side to the center cavity
(Fig.~1c), we obtain a band of bosonic modes which mimic two external waveguides. The corresponding
photon creation/annihilation operators in Eq.~(\ref{hamilton}) can in this limit be replaced by their
average (coherent state) values, $p_{1,3} \rightarrow \langle p_{1,3} \rangle=-2 i E_{1,3}/ \gamma$.  The
effective Hamiltonian of the system
\begin{equation}
\label{hamilton_mf}
      \hat{H} \sim \Delta_{2} \hat{p}_{2}^{\dagger}\hat{p}_{2}
      +U \hat{p}_2^{\dagger}\hat{p}_2^{\dagger}\hat{p}_2 \hat{p}_2
      +E_{\mathrm{eff}}\hat{p}^{\dagger}_2 + \mathrm{h.c.} \, ,
\end{equation}
then reduces to that of a single nonlinear cavity pumped by a coherent field with amplitude
$E_{\mathrm{eff}}=-2iJ(E_1 + E_3)/ \gamma$. We choose parameters such that $|E_{\mathrm{eff}}|=J$,
i.e. $J$ acts as the effective pumping rate. The steady state results for $g^{(2)}_2(0)$ and $\langle
n_2 \rangle$ are shown in Figs.~4a and 4b, respectively. The $g^{(2)}_2(0)$ displays a sharp transition
from Poissonian to sub-Poissonian light statistics as the interaction strength $U$ is increased. The
threshold for anti-bunched (sub-Poissonian) light generation, $U_{th}$, is a function of $J$. For
$J/\gamma_2 \ll 1$, the threshold is $U_{th} \sim \gamma_2$, in the opposite limit $U_{th} \sim J$.
At small hoppings, the nonlinearity  (i.e. anti-bunching) sets in when $U$ is larger than the
broadening of the bare center cavity polariton states i.e. $\gamma_2$. Much more interesting is the
fact that at larger hoppings the threshold scales with $J$, in contrast to the case of three cavities
(Fig.~2c). The crossover from bunching to antibunching behavior reflects in a clear way the crossover
from delocalized to localized states. As $J$ is increased, the relevant eigenstates of
the coupled system are superpositions of center and outer cavity states. When $J \gg U$ photon states
are delocalized over the whole systems while in the opposite case the states are Fock states due to
the onset of photon blockade. One may also understand the dependence of the crossover on $J$ by relating it to the low-$J$ limit of Fig.~2c: from the perspective of the center cavity, the driven waveguide is analogous to a driven cavity with a dissipation rate larger than all other energy scales.

Due to the Heisenberg uncertainty relation, the crossover from bunching
to antibunching behavior manifests itself also in the phase dependence of $\langle n_2 \rangle$ 
shown in Fig.~4b.
On increasing the interaction the visibility is strongly suppressed and furthermore there is a marked
deviation from the simple cosine law found for $U=0$. An analytical expression for the function in
the inset can be found in the infinite-$U$ limit, where Eq. (\ref{hamilton_mf}) is replaced by a
two-level system coupled with a driven cavity mode
\begin{equation}
\label{eqnn2largeU}
\langle n_2\rangle =\frac{\cos^2(\phi/2)}{2\cos^2(\phi/2)+(\gamma^2/8 J |E|)^2} \, ,
\end{equation}
which agrees with the numerical results.  The broadening of the dip at $\phi=\pi$ is proportional
to $\gamma^2/(4J|E|)$ implying that the visibility goes to zero by increasing 
the field ($\langle n_2(\phi=0)\rangle\to 0.5$). 
The behavior of $\langle n_2(\phi)\rangle$ is again witness of the crossover from the delocalized 
to the correlated regimes. For large hopping the state of the system is approximately a coherent state.
Phases are locked (there are strong fluctuations in the number operator) and the visibility is large.
On the opposite case, due to photon blockade the state is close to a Fock state. Phase fluctuations
in the central cavity suppress the global coherence of the system and the visibility is suppressed.
We note that although the functional forms are different, both $\langle n_2(\phi=0)\rangle$ and 
$g^{(2)}_2(0)$ display a cross-over that is dependent only on the dimensionless ratio $U/J$, 
as shown in Fig.~4c. 

\section{Summary}
In conclusion, we have studied the out-of-equilibrium interplay of tunneling and interactions in realistic quantum optical devices. For the three-cavity system, the inhibition of particle number fluctuations beyond Fock states $|0\rangle$ and $|1\rangle$ in the center-cavity reduces quantum coherence between the two outer ones. This is found to exhibit a threshold behavior as a function of the correlation energy $U$, which is clearly discernible from the saturation of Josephson oscillations at amplitude $\langle n_2 \rangle =0.5$, and by a strong deviation from cosine-like behavior at strong pumping. Remarkably, we find that the effective non-linearity of the system displays a cross-over into a JC-like non-linearity as $U/J$ is increased beyond $1$. Photon correlation measurements for this device reveal a sharp threshold from Poissonian to sub-Poissonian statistics that is almost insensitive to the strength of the tunnel-coupling $J$. On the other hand, in the case of a linear array of cavities coupled to a non-linear center cavity, the anti-bunching threshold is found to depend strongly on the tunnel-coupling. This observation signifies that photon correlation measurements are very effective in revealing the interplay of coherent tunneling and on-site interactions and may contain the key to interpret and probe possible phases of extended cavity-arrays which operate under non-equilibrium conditions.

\section{Methods}

The model in Eq. (\ref{hamilton}) is reformulated by introducing the canonically transformed bosonic operators
$\hat{s}=(\hat{p}_1+\hat{p}_3)/\sqrt{2}$ and $\hat{d}=(\hat{p}_1-\hat{p}_3)/\sqrt{2}$,
from which
\begin{equation}\label{hamilton2}
\hat{H}_{\mathrm{s}} =
\Delta (\hat{s}^{\dagger}\hat{s}  + \hat{p}_2^{\dagger}\hat{p}_2 ) +
\tilde{J}(\hat{p}_2^{\dagger}\hat{s} + \hat{s}^{\dagger}\hat{p}_2) +
U \hat{p}_2^{\dagger}\hat{p}_2^{\dagger}\hat{p}_2 \hat{p}_2  
+ \tilde{E}\hat{s}^{\dagger} + \tilde{E}^{\ast}\hat{s}  \, ,
\end{equation}
where we defined $\Delta_k=\Delta$ and
we discarded the dynamics of the field $\hat{d}$, which is decoupled from
$\hat{p}_2$. Rescaled quantities are defined as
$\tilde{J}=\sqrt{2}J$ and $\tilde{E}=\sqrt{2}(E_1 + E_3)/2$. Thus, the dynamics of the
full model (\ref{hamilton}) is equivalent to that of two coupled bosonic fields: $\hat{s}$ is
coherently driven, while $\hat{p}_2$ is nonlinear.
Losses are taken into account within the
quantum master equation in Born-Markov approximation for the system density matrix,
Eq. (\ref{master}), for the field operators $\hat{s}$ and $\hat{p}_2$ with dissipations $\gamma$ and
$\gamma_2$, respectively.

\textit{Analytical solution}.
An analytical solution to the steady state master equation can be found in the non-interacting limit,
$U=0$.
The equation of motion for a generic operator expectation value, $\langle \hat{A}\rangle$, is $\partial \langle \hat{A} \rangle / \partial t =0=i \langle [\hat{H}_{\mathrm{s}}, \hat{A}] \rangle_{\mathrm{ss}} +
\langle \mathcal{L} [\hat{A}] \rangle_{\mathrm{ss}}$, where $\hat{H}_{\mathrm{s}}=(\tilde{J} \hat{p}_2  + \tilde{E} ) \hat{s}^{\dagger} + \mathrm{h.c.}$, and the Liouvillian $\mathcal{L} [\hat{A}]$ is formally written as
\begin{equation}
\label{liouvil}
     \mathcal{L} [\hat{A}] =
     \frac{\gamma}{2}
      (2\hat{s}^{\dagger} \hat{A} \hat{s} - \hat{s}^{\dagger}\hat{s} \hat{A}
      - \hat{A} \hat{s}^{\dagger}\hat{s})
    + \frac{\gamma_2}{2}
      (2\hat{p}_2^{\dagger} \hat{A} \hat{p}_2-\hat{p}_2^{\dagger}\hat{p}_2 \hat{A}
      - \hat{A} \hat{p}_2^{\dagger}\hat{p}_2)\, .
\end{equation}
Solving for $\hat{p_2}$ and $\hat{s}$, respectively, we get a system of two coupled equations, from which the steady state solution is
$|\langle \hat{p_2} \rangle_{\mathrm{ss}}|=|\tilde{E}|/\{\tilde{J} [(1+\gamma\gamma_2/(4\tilde{J}^2)]\}$,
and hence Eq. (\ref{solutionUeq0}).

\textit{Numerical solution}.
Extensive numerical simulations for the effective model (\ref{hamilton2}) can be performed
quite efficiently and in a reduced Hilbert space with respect to the full model.
After explicitly defining the operators in matrix form on a Fock basis of bosonic number states,
the steady state density matrix for any given set of parameters can be obtained by finding
the eigenvector corresponding to the zero eigenvalue of the linear operator equation
$\hat{L}|\rho\rangle\rangle=\lambda|\rho\rangle\rangle$, where
$|\rho\rangle\rangle$ is the density operator mapped into vectorial form, and $\hat{L}$
is the linear matrix corresponding to the Liouvillian operator in the right-hand side of Eq.
(\ref{master}) \cite{stenholm03pra,zoller08}.
Once $|\rho_{\mathrm{ss}}\rangle\rangle$ is obtained from
$\hat{L}|\rho\rangle\rangle_{\mathrm{ss}}=\lambda_{\mathrm{ss}}
|\rho\rangle\rangle_{\mathrm{ss}}$ with $\lambda_{\mathrm{ss}}=0$,
we can recast it in matrix form and calculate any observable we
are interested in. In particular, in this work we deal with
$\langle n_{2} \rangle=Tr\{\hat{p}_{2}^{\dagger}\hat{p}_{2}\rho_{\mathrm{ss}}\}$, and
the steady state zero-time delay second-order correlation function
$g^{(2)}_{2}(\tau=0)=Tr\{\hat{p}_{2}^{\dagger}\hat{p}_{2}^{\dagger}\hat{p}_{2}
\hat{p}_{2}\rho_{\mathrm{ss}}\}/\langle n_{2}\rangle^2$.
To check convergence with the number of Fock states in the basis as a function of
$\tilde{J}$, numerical results for $U=0$ are compared to Eq. (\ref{solutionUeq0}).

\section{Supplementary Information}

\textit{Experimental feasibility}.
In the main text we have considered a generic Kerr nonlinearity as the source of strong 
photon correlation in the center cavity.
A possible way of experimentally implementing an effective hamiltonian of the type
(1) or (5) in the main text is to couple 4-level atomic ensembles with microtoroid 
resonators, as described in detail in the literature \cite{hartmann06}. 
The latter certainly represents an interesting
possibility for a practical realization of our proposal with state-of-the 
art atomic cavity QED.
Our focus here will be on the scheme represented in Fig. 1b and d (see text), in which
a quantum dot (QD) is assumed to be deterministically coupled to the high-Q 
photonic crystal cavity mode in the middle \cite{kevin07nat}. It has been experimentally
shown that such a system displays
single-photon nonlinearities under coherent resonant pumping \cite{jelena08}. 
Photonic crystal (PC) circuits allow for a straightforward on-chip 
implementation with side-coupled cavities (as in Fig. 1b) or waveguides (as in Fig. 1d). 
If the hopping parameter $J$  is small compared to the laser intensities $|E_1|$ and 
$|E_3|$, we can approximate the states of the external cavities with  coherent fields of 
intensity $2 |E_{1,3}|/ \gamma$ (with $\gamma$ being the damping parameter of the 
external cavities).
Thus, to first order in $J/|E_{1,3}|$, the dynamics of the central cavity can be effectively 
described by replacing in the Hamiltonian the operators
$\hat{p}_1$ and $\hat{p}_3$ with  $2 |E_{1,3}|/ \gamma$, 
with reference to the derivation of the model in Eq. (5) in the text. 
With this choice the 3-cavity set-up can be effectively reduced to
a Jaynes-Cummings model coupled to external driving fields
and described by the simplified model
\begin{equation}
\label{hamilton_mf_jc}
      \hat{H} \sim \Delta_{c} \hat{a}_{2}^{\dagger}\hat{a}_{2}+
      \Delta_{x} \hat{\sigma}_{+}\hat{\sigma}_{-}
      + i g (\hat{a}_2^{\dagger}\hat{\sigma}_{-} - \hat{\sigma}_{+}\hat{a}_2 )
      +E_{\mathrm{eff}} \hat{a}^{\dagger}_2 + E_{\mathrm{eff}}^{\ast} \hat{a}_2\, ,
\end{equation}
where $\hat{a}_{2}$ ($\hat{a}_{2}^{\dagger}$) represents annihilation (creation) of cavity
photons, while  $\hat{\sigma}_{-}$, and $\hat{\sigma}_{+}$ are 
Pauli lowering and rising operators related to the effective two-level system
representing the QD exciton transition; 
$g=\hbar (\pi e^2 f/ \varepsilon m^{\ast} V_{\mathrm{eff}})^{1/2}$ 
is the exciton-photon coupling (expressed
in  terms of the effective cavity mode volume and the QD oscillator strength) \cite{andreani99prb},
and $\Delta_c=\omega_c-\omega_{\mathrm{L}}$ and $\Delta_x=\omega_x-\omega_{\mathrm{L}}$ 
are the cavity and exciton detuning from the pump frequency, respectively. The effective pumping
strength is $|E_{\mathrm{eff}}| \propto |E_1 + E_2 | / \gamma$ and is responsible for  
the effects of phase detuning, $\phi = \phi_1 - \phi_3$.

To give some numbers, state-of-the art solid state QED with GaAs-based materials allows for 
$Q\sim 10^5 - 10^6$, i.e. realistic $\gamma_c \sim 0.01$ meV \cite{derossi08apl},
$\gamma_x\sim 1.3 \times 10^{-3}$ meV \cite{mete07nl}, 
$g=0.11$ meV \cite{kevin07nat}.
In the experiment, $g$ is fixed as well as $J$, but the effective single-photon nonlinearity
can be tuned by changing $\delta=\omega_x-\omega_c$. There are a number of different
techniques to deterministically tune the cavity mode 
frequency \cite{kevin07nat,badolato05sci,strauf06apl,kevin06apl} and/or the QD exciton 
resonance \cite{rastelli}. 
For the system excitation, the same laser source can be sent through a 
beam-splitter, one of the arms going directly into the PC circuit (e.g. through a tapered 
access waveguide) with phase $\phi_1$, while the other being delayed and sent 
through a second tapered waveguide into the circuit with phase $\phi_3$.
Given that lasers with sub-MHz linewidth (i.e. much smaller than cavity and exciton
dissipation rates) are currently available, 
we do not regard possible phase fluctuations in the two driving fields as a limiting issue 
for this scheme to be realized.
Finally, it is key to this experiment that the pump laser 
frequencies be tuned to the lower polariton frequency of the JC spectrum, i.e. for 
each detuning $\delta$ we set $\omega_{\mathrm{L}}\simeq (\omega_x+\omega_c)/2 -
\sqrt{g^2 + \delta^2 / 4}$ \cite{kevin07nat}. 

\begin{figure}[t]
\begin{center}
\includegraphics[width=\linewidth]{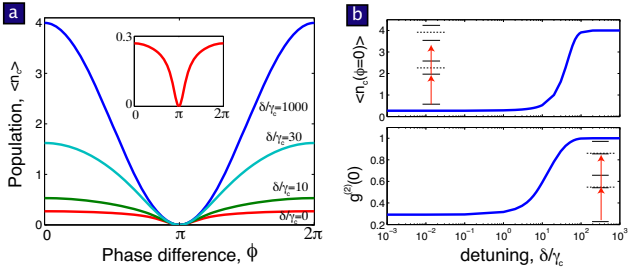}
\caption{Numerical simulation of the experimental feasibility. 
We assume parameters of the model in Eq. (\ref{hamilton_mf_jc}) 
$g/ \gamma_c = 10$, $g/ \gamma_x = 100$ (as realistically achievable, see text), 
and effective pumping strength $|E_{\mathrm{eff}}|/ \gamma_c =1$ at $\phi=0$. 
Results are shown for light intensity 
and second-order correlation function emitted from the middle cavity, i.e. 
$\langle n_c \rangle=\langle \hat{a}_{2}^{\dagger}\hat{a}_{2} \rangle$ and
$g^{(2)}(0)=\langle \hat{a}_{2}^{\dagger}\hat{a}_{2}^{\dagger}\hat{a}_{2}\hat{a}_{2} \rangle
/ \langle n_c \rangle^2$. (a) Josephson-like oscillations are suppressed when the cavity-exciton
detuning $\delta\to 0$. When $\delta\gg \gamma_c ,g $, i.e. the exciton resonance is strongly
blue-detuned from the cavity mode, the lower polariton is more and more cavity-like, hence the
effective nonlinearity of the system is tuned  through $\delta$. In the inset, a zoom on the
curve for $\delta=0$. (b) The crossover from correlated
to delocalized regimes is shown for $\langle n_c \rangle$ and $g^{(2)}(0)$ at $\phi=0$, 
respectively,  as a function of $\delta$.  } \label{fig1SI}
\end{center}
\end{figure}

\begin{figure}[t]
\begin{center}
\includegraphics[width=\linewidth]{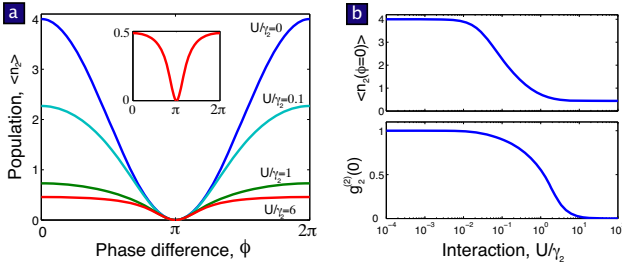}
\caption{Numerical simulation for the model of Eq. (5) in the main
text with effective pumping strength $|E_{\mathrm{eff}}|/ \gamma =1$ at $\phi=0$. 
(a) Josephson-like oscillations are suppressed on increasing the effective Kerr nonlinearity, 
$U/ \gamma_2 $. In the inset, a zoom on the
curve for $U/ \gamma_2 = 6$, to be compared to the previous figure. 
(b) Crossover from correlated
to delocalized regimes for $\langle n_2 \rangle$ and $g_2^{(2)}(0)$ at $\phi=0$, 
respectively,  as a function of $U/ \gamma_2$.  } \label{fig2SI}
\end{center}
\end{figure}

We simulate this model by solving the corresponding master equation that incorporates
realistic cavity and exciton dissipation rates. Figure  \ref{fig1SI} shows that 
the predictions of the Jaynes-Cummings model in Eq. (\ref{hamilton_mf_jc})
are qualitatively similar to
the ideal Kerr nonlinearity model of Eq. (5) in the main text.  
In particular,  the dependence of light intensity on $\phi$ shows suppression of 
Josephson-like oscillations as the exciton frequency is tuned (from the blue-side)
in resonance with the cavity mode. 
The increase of oscillations amplitude towards the
bare-cavity limit appears together with a crossover from sub-Poissonian 
to Poissonian statistics as a function of $\delta$, consistent with the results of Fig. 4 in the 
main text.
Here, tuning $\delta$ is a way of effectively tuning the nonlinearity in 
the central cavity, and thereby to experimentally observe the 
crossover from tunnel-coupled to strong correlated photon dynamics in a state-of-the 
art device. Notice that the linear regime is completely recovered for $\delta/ \gamma_c \simeq 100$,
i.e. $\delta\sim 1$ meV with the parameters given above, which is perfectly within reach of present
experimental capabilities. 
In order to compare the behaviour of the different models,
we notice that the strongly correlated limit of Fig.~\ref{fig1SI}a ($\delta=0$, curve in the inset) 
is qualitatively similar to the result obtained with a Kerr nonlinear model, Eq. 
(5) in the text, for $U=(2-\sqrt{2})g$, i.e. $U/ \gamma_2 \simeq 6$, as it is 
shown in Fig.~\ref{fig2SI}a and its inset. The crossover from delocalized to localized
regimes as a function of $U$ is shown in Fig.~\ref{fig2SI}b to be compared to 
Fig.~\ref{fig1SI}b.
The strict analogies between the experimental realization proposed 
and the model considered in the main text before is maintained as long as 
the pumping strength of the Jaynes-Cummings system is low enough to discard
higher-lying photon manifolds. 

Finally, we stress that the same model and experimental approach can be used to study 
the realization of quantum optical Josephson interferometers based on different 
technologies, such as circuit QED \cite{schuster07nat}.
Moreover, we point out again that the effective Kerr nonlinearity can be achieved,
in the same system represented in Figs. 1b and d in the text, with a QW in strong coupling to 
the cavity mode. Although there is no experimental evidence of Kerr nonlinear behaviour
of  3D confined cavity polaritons at time of writing, it is likely that polariton blockade on a 
single quantum box can be achieved in the near future along the lines and numbers quoted 
in the literature \cite{ciuti06prb}, e.g. with an alternative cavity geometry recently 
realized \cite{eldaif06apl}. Such a result would make possible a more direct realization 
of our model with a Kerr nonlinearity in the solid state.

\begin{figure}[t]
\begin{center}
\includegraphics[clip,width=\linewidth]{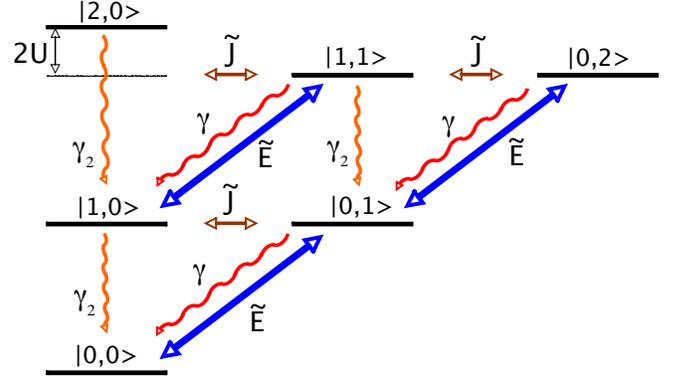}
\caption{Energy level diagram and rates for the coupled cavity system. The various $N_{tot}$ manifolds are set off from each other by an arbitrary $\omega_p$ for visibility. The numbering of the states $\bra{n}$ are from down to up and left to right. For instance $\bra{1} = \bra{1,0}$.}
\label{fig3SI}
\end{center}
\end{figure}

\textit{Derivation of Eq. (4)}. Here we investigate the steady state dynamics of the Hamiltonian (7) (see Methods section, main text) in the weak pumping limit. We will present a derivation of the  expressions for $\langle n_2(\phi=0)\rangle$ and $g^{(2)}_{2}(\tau=0)=\langle\hat{p}_{2}^{\dagger} \hat{p}_{2}^{\dagger}\hat{p}_{2}\hat{p}_{2}\rangle/\langle n_{2}\rangle^2$. Consider the low-energy excitations of the Hamiltonian (7) in the basis $\bra{n_2,n_s}$ where $N_{tot}=n_2 + n_s$ is the total number of photons. In the weak pumping limit, the drive term $\tilde{E}\hat{s}^{\dagger} + \tilde{E}^{\ast}\hat{s}$ causes transitions between the manifolds $N_{tot}$ and $N_{tot}+1$. Let us write the total time-dependent wavefunction of the system as $\Psi(t) = \sum_n a_n \bra{n}$. In the weak pumping limit, we consider the lowest three total photon manifolds $N_{tot} = 0, 1, 2$, hence $n=0,\ldots,6$. The corresponding energy level diagram and the rates are shown in Fig.~\ref{fig3SI}. 

Terms neglected in $\Psi(t)$ will be of order $O(\frac{|\tilde{E}|}{\bar{\gamma}})^3$ where $\bar{\gamma}$ is the typical decay rate of the system ($\gamma_2$, $\gamma$). It's important to write the equations of motion in the bare basis instead of the basis of the coupled cavity states to get the dissipation rates correctly. The equations of motion are
\begin{eqnarray*}
\dot{\tilde{a}}_0 & = & 0 \\
\dot{\tilde{a}}_1 & = & -\left( i\Delta + \frac{\gamma_2}{2} \right)  \tilde{a}_1 - i\tilde{J} \tilde{a}_2 \\
\dot{\tilde{a}}_2 & = & -\left( i\Delta + \frac{\gamma}{2} \right)  \tilde{a}_2 - i\tilde{J} \tilde{a}_1 - i\tilde{E}\tilde{a}_0 \\
\dot{\tilde{a}}_3 & = & -\left( 2i\Delta + 2iU + \gamma_2 \right)  \tilde{a}_3 - i\sqrt{2} \tilde{J} \tilde{a}_4 \\
\dot{\tilde{a}}_4 & = & -\left( 2i\Delta + \frac{\Gamma}{2} \right)  \tilde{a}_4 - i\sqrt{2} \tilde{J} (\tilde{a}_3 +  \tilde{a}_5) - i \tilde{E}\tilde{a}_1 \\
\dot{\tilde{a}}_5 & = & -\left( 2i\Delta + \gamma \right)  \tilde{a}_5 - i\sqrt{2} \tilde{J} \tilde{a}_4 -  i \tilde{E}\tilde{a}_2
\end{eqnarray*}
We have kept terms that are the same order of magnitude in $\frac{|\tilde{E}|}{\bar{\gamma}}$ and $\tilde{a}_n = a_n \ex{i\omega_{\mathrm{L}} t}$ \cite{fn1}. The steady-state solutions can be easily determined by additionally employing the normalization condition $\sum_{n=1}^6 |a_n|^2 = 1$ (to order $O(\frac{|\tilde{E}|}{\bar{\gamma}})^6$). We find, for $\Delta = 0$, 
\begin{equation}
a_1 = - \frac{4\tilde{J}\tilde{E}}{\gamma\gamma_2 + 4\tilde{J}^2}, \quad \quad a_2 = - \frac{2i\tilde{E} \gamma_2}{\gamma\gamma_2 + 4\tilde{J}^2}
\end{equation}
It's interesting to note that in the limit $J/\gamma \gg 1$, $|a_1| \gg |a_2|$, showing that interference effects play an important role. Thus, 
\begin{equation}
n_p \sim |a_1|^2 = 16\tilde{J}^2\tilde{E}^2/(\gamma\gamma_2 + 4\tilde{J}^2)^2
\end{equation}
to order $O(\frac{|\tilde{E}|}{\bar{\gamma}})^4$. Solving for the two-photon manifold amplitudes as well, we find (4)
\begin{equation}
g^{(2)}_{2}(\tau=0) \sim \frac{2|a_3|^2}{|a_1|^4} =  \frac{\Gamma^2}{\Gamma^2 + 
4\alpha^2(\tilde{J})U^2}
\end{equation}

\begin{acknowledgements}
 The authors would like to acknowledge useful discussions with I. Carusotto, C. Ciuti, and S. De Liberato. This work was partly supported by NCCR Quantum Photonics. R.F. acknowledges financial support from EUROSQIP.
Correspondence and requests for materials should be addressed to D.G.~(email: gerace@fisicavolta.unipv.it).
\end{acknowledgements}


\begin{thebibliography}{99}
\bibitem{fn1}
Te Hamiltonian in Equation~(7) in the main text is written in the interaction picture. Here the wavefunction $\Psi$ is the wavefunction written in the Schr\"odinger picture for completeness.
\end{thebibliography}


\end{document}